# Ion-beam assisted laser fabrication of sensing plasmonic nanostructures

Aleksandr Kuchmizhak,[*] Stanislav Gurbatov, Oleg Vitrik, Yuri Kulchin, Valentin Milichko, Sergey Makarov, and Sergey Kudryashov

Aleksandr Kuchmizhak, Stanislav Gurbatov, Oleg Vitrik

*School of Natural Sciences, Far Eastern Federal University, 8 Sukhanova str., Vladivostok 690041, Russia*

*Institute of Automation and Control Processes, Far Eastern Branch, Russian Academy of Science, 5 Radio str., Vladivostok 690041, Russia*

Yuri Kulchin

*Institute of Automation and Control Processes, Far Eastern Branch, Russian Academy of Science, 5 Radio str., Vladivostok 690041, Russia*

Valentin Milichko, Sergey Makarov

*ITMO University, St. Petersburg 197101, Russia*

[¶]Sergey Kudryashov

*ITMO University, St. Petersburg 197101, Russia*

*Lebedev Physical Institute, Russian Academy of Science, Moscow 119991, Russia*

Corresponding author : Aleksandr Kuchmizhak (e-mail: alex.iacp.dvo@mail.ru)

**Abstract.** Simple high-performance two-stage hybrid technique was developed for fabrication of different plasmonic nanostructures, including nanorods, nanorings, as well as more complex structures on glass substrates. In this technique a thin noble metal film on a dielectric substrate is irradiated by a tightly focused single nanosecond laser pulse and then the modified region is slowly polished by an accelerated argon ion ($Ar^+$) beam. As a result, each nanosecond laser pulse locally modifies the initial metal film through initiation of fast melting and subsequent hydrodynamic processes, while the following $Ar^+$-ion polishing removes the rest of the film, revealing the hidden topography features and fabricating separate plasmonic structures on the glass substrate. We demonstrate that the shape and lateral size of the resulting functional plasmonic nanostructures depends on the laser pulse energy and metal film thickness, while subsequent $Ar^+$-ion polishing enables to vary height of the resulting nanostructures. The plasmonic properties of the fabricated nanostructures were characterized by dark-field micro-spectroscopy, Raman and photoluminescence measurements from single nanofeatures, as well as by supporting numerical calculations of the related electromagnetic near-fields and Purcell factors. The developed simple two-stage technique represents a next step towards direct large-scale laser-induced fabrication of highly-ordered arrays of complex plasmonic nanostructures.

Functional plasmonic nanostructures (FPNs) fabricated on transparent dielectric substrates as ordered arrays of nanofeatures are the object of increasingly growing scientific interest in the last decade. Such structures exhibit unique electromagnetic properties allowing manipulation and amplification of incoming visible light at nanoscale through excitation of coherent electron plasma oscillations referred to as localized plasmon resonance (LPR)[1]. Many promising applications of LPR in such rapidly developing fields of science as biosensing, photovoltaics, nanophotonics, near-field optical microscopy and nanomedicine were theoretically predicted and successfully demonstrated experimentally in the past two decades[2-8]. Modern scientific problems, arising in these scientific areas, impose stringent requirements on FPN fabrication techniques, which must enable control over geometric dimensions of each individual FPN feature in their ordered arrays, combined with high overall functional performance and repeatability, as well as relatively low cost. The opportunity to fabricate a single nanostructure at a given point on a sample surface, for example, on a cut edge of an optical fiber or a scanning probe tip is also desirable. In this lieu, advanced bottom-up approaches, utilizing direct impact of a focused accelerated ion beam or a combination of electron milling and subsequent post-processing steps, enable fabrication of separate FPNs or their small arrays with very high precision down to the angstrom level. However, these approaches become extremely expensive and time-consuming, when the number of FPN features in their arrays approaches $10^4$-$10^6$. Other well-established nanofabrication techniques as chemical synthesis [9-11], colloidal lithography [12], laser- or ion-beam-assisted dewetting of thin metal films [13,14] are suitable for large-scale fabrication of only disordered FPN arrays, providing high uniformity of geometric shapes and spatial dimensions of their constituent elementary nanofeatures, but require additional pre-processing fabrication steps to arrange isolated elements into highly ordered arrays [15-18].

In the last decade, direct laser nanostructuring of metal films by short (nanosecond) and ultrashort (femtosecond) laser pulses was demonstrated to be a promising alternative to ion-beam milling and chemical synthesis techniques. The impact of such pulses initiates a fast phase transition of a solid metal film into a molten layer, with its subsequent dynamics prior recrystallization defined by a sequence of well-established thermal and hydrodynamic processes. Spatial scale and final topography of these processes, determining the shape of the resulting nanostructures, is controlled by the metal film irradiation conditions (pulse energy and duration, focal spot size etc.), as well as by the physical properties of the metal film and its substrate (chemical composition, thickness, optical and thermal characteristics etc.). In particular, such approach was proved to be an efficient route for direct single-pulse fabrication of ordered arrays of resolidified nanojets (upright standing metal tips [19-20]), which are suitable for LPR-mediated amplification of electromagnetic fields in near-IR spectral region [21] and polarization-dependent Raman signal enhancement [22], imprinting of two-dimensional coupling elements through the excitation of intense surface plasmon-polariton (SPP) waves [23], as well as for laser-induced forward transfer (LIFT) of spherical metal and semiconductor droplets onto an acceptor substrate [24,25].

Meanwhile, up to date capability of such direct laser nanostructuring techniques was limited by fabrication of relatively simple FPNs - spherical nanoparticles, 2D surface gratings or nanojets, leaving the fabricated structures attached to the initial pre-deposited metal film (with the only exception for the LIFT method). Also, rather expensive femtosecond lasers are often used for this purpose [19-25]. In this paper, we demonstrate for the first time that the use of direct single-shot exposure of the metal film by tightly focused nanosecond (ns) laser pulses, followed by slow polishing of the fabricated nanostructures by an accelerated argon ion ($Ar^+$) beam, in combination allows fabricating on glass substrates arrays of isolated plasmonic nanorods, separated and merged nanorings, as well as more complex nanostructures. Within this approach, the ns-laser irradiation of metal film changes its initial thickness through the initiation of fast melting and subsequent hydrodynamic processes, while the following $Ar^+$ polishing reveals the features of its hidden topography, producing LPR-supporting isolated plasmonic structures on a glass substrate. We experimentally demonstrate that both the shape and the lateral size of the resulting FPNs are determined by ns-laser pulse energy and metal film thickness, while the subsequent $Ar^+$-ion polishing allows varying the height of the resulting nanostructures. Their plasmonic properties were examined by means of dark-field (DF) micro-spectroscopy, micro-Raman and photoluminescent (PL) measurements and were modeled in the framework of finite-difference time-domain calculations of electromagnetic near-fields and related Purcell factor (PF) magnitudes.

**Results**

Nanostructuring of a "thermally thin" noble metal film on a bulk glass substrate by a single ns-laser pulse with a Gaussian-like lateral energy distribution can be formally divided as a function of laser pulse energy into five main steps/regimes (Fig.1a-e), where each of these regimes for a certain film thickness d is characterized by its threshold pulse energy $E_{th,i}$ (i is the regime number). Specifically, the first step (Fig. 1a, $E>E_{th1}$) is related to melting and subsequent detaching of a metal film from its substrate due to thermal stresses caused by rapid lateral expansion of the metal film[26], and/or via evaporation process at the "film-substrate" interface[27], yielding in a microbump.

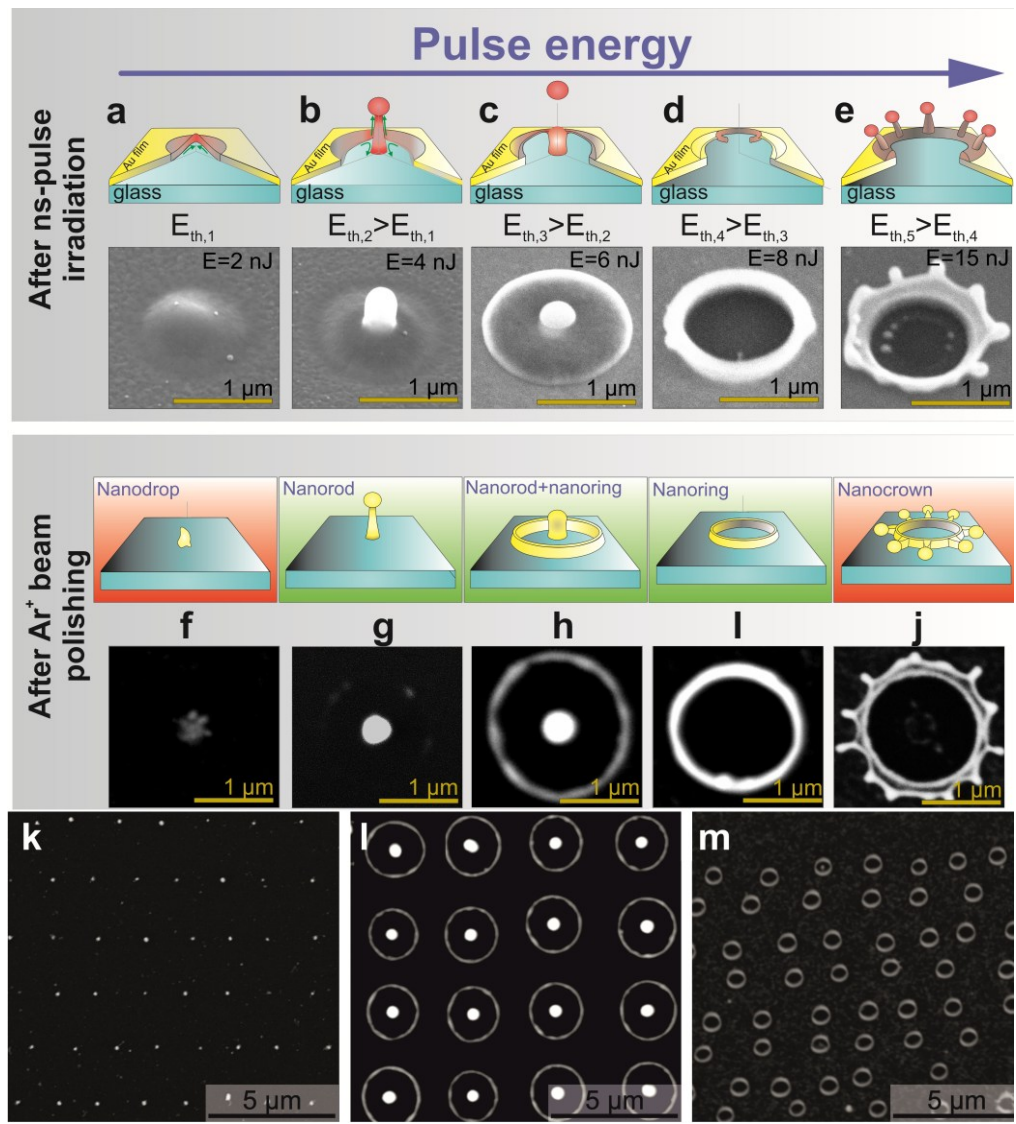

**Figure 1. Two-stage fabrication of functional plasmonic nanostructures.** (a-e) Schematic illustrations of consequent evolution steps during nanostructuring of a noble metal film on a glass substrate irradiated by a single tightly focused nanosecond pulse at different pulse energies E. Each step is illustrated by the corresponding bottom SEM image of the resolidified nanostructure on the 75-nm-thich Au film surface upon its irradiation by a single tightly focused 7-ns laser pulse (the pulse energy E increases from left to right); (f-j) Schematic illustrations and corresponding SEM images of functional plasmonic nanostructures obtained after Ar$^+$-ion beam polishing of the abovementioned structures; (k-m) SEM images of regular arrays of 300-nm wide, 500-nm high nanorods, of larger nanorods surrounded by microrings, and of 600-nm wide nanorings, demonstrating the reproducibility of the two-stage fabrication technique.

The next step (Fig. 1b, E>$E_{th2}$) is characterized by an onset of strong thermocapillary forces and corresponding microscopic hydrodynamic flows (e.g., Marangoni convectional flow[28]), or a point-like phase explosion immediately after exceeding the corresponding threshold energy,[29] which pulls the molten film up and toward the optical spot center. As a result, non-local action of these forces causes the formation of the so-called nanojet in the center of the microbump, which height increases versus the laser pulse energy E[19,20,30]. These two steps demonstrate almost identical character under the action of either nanosecond, or femtosecond laser pulses with the same initial optical spots, while the main difference consists in the thicker lateral size (diameter) and lower height of the nanojets observed under the ns-laser irradiation.

The third step (Fig. 1c, E> $E_{th3}$) is characterized by two parallel processes. First, when the height of the liquid nanojet reaches some critical size, the onset of Rayleigh-Plateau hydrodynamic instability[13,31] breaks up its upper end, resulting in ejection of a spherical droplet. Second, significant blowing up of the microbump, surrounding the nanojet, is observed apparently owing to increase of the vapor pressure in the cavity formed at the "film - substrate" interface[27]. This effect becomes effective only at ns-laser irradiation of relatively "thick" films (particularly, ones thicker, than 50 nm), supporting the overheating of the metal melt accompanied by lateral phase explosion[27,29].

At further increasing of the pulse energy (E>$E_{th4}$, Fig.1d), the dramatically (usually, almost by 10 times[32, 33]) thinned microbump cladding breaks up, forming a through micro-hole with a pronounced molten rim. The diameter of the micro-holes increases with the pulse energy E up to the certain critical size determined, as shown below, by the film thickness d. Upon reaching this critical size, the onset of some hydrodynamic instability (Rayleigh-Plateau, Marangoni ones[21]) results in periodic modulation of the height of the molten rim walls and subsequent formation of the so-called nanocrown structure[34,35].

Each of the abovementioned steps is characterized by the local change/redistribution of the initial film thickness on the flat substrate surface, i.e. concentrating or thinning, and, finally, resolidifying the molten film at some specific positions across the focal spot. Such modified surface relief has not only its visible upper topography, but also its hidden rear-surface part, which can be revealed by removing the rest part of the metal film using methods of chemical etching (for example, in $HNO_3$ acid for Ag, or in $HNO_3$ + 3HCl acid mixture for Au), similarly to the process of photoresist treatment after its irradiation by an electron beam. However, use of chemical reagents usually leads to surface contamination and subsequent post-processing cleaning to remove the etching residues. In this work we used for such etching slow polishing of the sample by an unfocused accelerated $Ar^+$-ion beam, with the results of this procedure shown in (Fig.1f-j) for each nanostructuring step. As seen, the impact of $Ar^+$-ion beam leads to formation of different isolated nanostructures at specific places on the sample surface previously irradiated by a single ns-laser pulse: nanorods, nanorings, as well as nanorod-nanoring ensembles. The polishing procedure made for microbumps (Figs.1a,f) results in formation of nanostructures with irregular geometrical shapes and positions relative to the laser spot center, as the resolidified molten material concentrated mainly at the microbump top is not connected to the substrate. Moreover, in the case of nanocrowns (Figs.1e,j) the number and the relative position of the circumferentially spaced nanotips can significantly vary from structure to structure. In other cases (nanorings and nanorod-nanoring ensembles), the shapes of corresponding isolated nanostructures demonstrate reproducibility, sufficient for fabrication of different large-scale FPN arrays (see Figs.1k-m).

For the most reproducible fabrication regimes (Fig.1h,i), when the formation of ring-shaped and nanorod-nanoring ensembles are observed, the proposed two-stage nanofabrication technique was systematically tested on its robustness and possibility of tailoring of main geometric parameters of the fabricated nanostructures, dictating their spectral LPR position as well as their potential application areas[36-38].

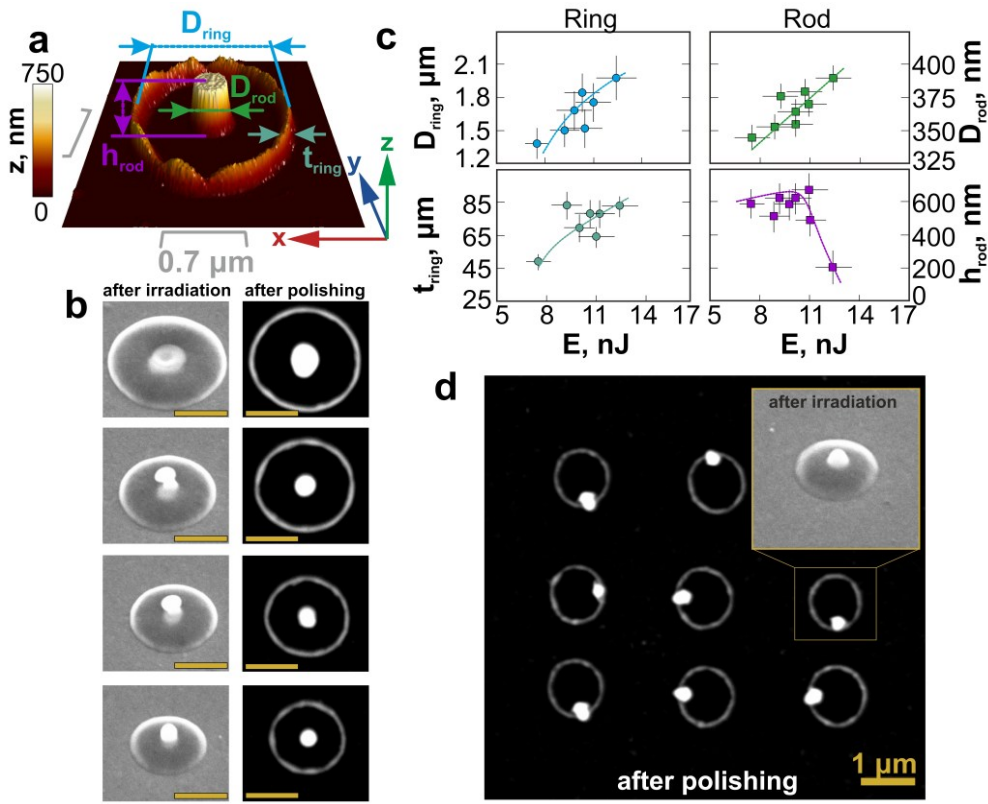

**Figure 2. Fabrication and tailoring dimensional parameters of nanoring-nanorods ensembles.** (a) AFM image of a nanorod-nanoring ensemble on a glass substrate, indicating its main geometric parameters; (b) two series of SEM images of the nanorod-nanoring ensemble after the single ns-pulse

irradiation (left row) and after subsequent Ar$^+$-ion beam polishing (right row), illustrating their shape evolution with the increase of the laser energy E (E increases from the bottom to the top); (c) dependences of main geometric parameters of nanorod-nanoring ensembles obtained for the 75-nm thick Ag-film on E; (d) SEM image of the square 3x3-array of the nanorod-nanoring ensembles fabricated in the 35-nm thick Au-film. The inset shows the side-view (angle of 45°) SEM image of the single nanofeature prior its Ar$^+$-ion beam polishing.

Fabrication of a set of a nanorod surrounded by a nanoring (referred here to as a "nanorod-nanoring ensemble") is based on the post-processing by Ar$^+$-ion polishing removal of the significantly thinned microbump cupola[32] between the central nanojet and the surrounding molten rim, which were both formed after resolidification at the energies higher, than the threshold value $E_{th2}$. Moreover, in this case variation of the pulse energy enables tuning of the main dimensional parameters (Fig. 2a) of the nanorod-nanoring ensemble: the height $h_{rod}$ and the diameter $D_{rod}$ of the central nanorod, as well as the diameter $D_{ring}$, the height $h_{ring}$ and the thickness $t_{ring}$ of the nanoring wall. The next series of SEM images (the left row in Fig. 2b) illustrates the tendency of microbump diameter to increase at increasing the pulse energy E. The subsequent ion polishing of such nanostructures (the right row in Fig. 2b) reveals an axially symmetric nanorod-nanoring ensemble, which contains an isolated central nanorod within a nanoring. The nanoring height $h_{ring}$ is equal to the initial film thickness d, though slightly varies owing to the uneven thickness of the microbump walls. The average diameter of the nanorod $D_{rod}$ also increases versus E, while the nanorod height $h_{rod}$ slightly increases with the increasing pulse energy, resulting from more intense melt flow in the microbump center at the higher pulse energies, but prior formation and ejection of a spherical droplet (Fig. 2c). The nanorod height $h_{rod}$ usually exceeds the height $h_{ring}$ of the surrounding nanoring, providing the opportunity to fabricate isolated disk-like ($h_{rod}/D_{rod}<1$) and rod-like ($h_{rod}/D_{rod}<1$) nanostructures (Fig.1k). Importantly, the formation process of such nanorod-nanoring ensembles requires highly symmetrical energy distribution in the laser focal spot as well as the high adhesion of the central nanojet to the glass substrate, which is realized through a counter-jet formation mechanism[28] (Fig. 1b). Such adhesion of the nanojet with the glass substrate appears, when sufficiently large amount of the molten material, resulting from the large initial metal film thickness, takes part in the nanojet formation. For this reason, the formation and subsequent polishing of the nanostructures fabricated in the relatively thin metal films (d < 50 nm) results in random position of their central nanojets with respect to the nanoring centers (Fig. 2d), owing to the weak adhesion of the nanojet to the glass substrate.

For simple nanorings their main parameters are diameter $D_{ring}$ (Fig. 3a), thickness t and height h of their walls[36-40]. As mentioned above, the diameter $D_{ring}$, which is directly determined by the size of the through-hole formed via the microbump ablative destruction, increases versus E (Fig. 3b). Such general tendency is observed for all film thicknesses, ranging from 15 to 120 nm, as shown in Fig. 3c as a set of curves in the coordinates $D^2_{ring}$ - lnE [41] for $E>E_{th3}$. The linear slope of these $D^2_{ring}(lnE)$ dependences, which determines the characteristic Gaussian diameter $D_{1/e}$ of the corresponding surface energy density distribution, increases versus film thickness d from 1.1 μm at d = 15 nm to 2.7 μm at d = 120 nm, in good agreement with the previously published results[42,43]. The threshold pulse energy $E_{th}$ required for such through-hole formation is also naturally increasing versus d. The upper part in Fig. 3c marked with the grey color, shows the range of laser pulse energies, where the formation of nanocrowns is observed for each film thickness. This range defines the maximum and minimum possible diameters $D_{ring}$ of the fabricated nanorings for each value d, which both gradually increase versus d. These relationships for the minimal and maximum diameters are presented in Fig. 3d for various film thicknesses, indicating that the smallest nanoring diameter ($D_{ring}$ ~0.55 μm) can be achieved for the thinnest 15-nm-thick film, while the largest one ($D_{ring}$ ~3 μm) corresponds to the 120-nm thick film. The surface of the nanorings demonstrates even finer roughness, comparing to the roughness of the initial nanocrystalline metal film on the glass substrate (see inset in Fig. 3c), owing to the liquid-state process of the nanoring formation during the ns-laser irradiation. Moreover, the $Ar^+$-ion beam polishing procedure, which completely removes the rest of the unmodified metal film, doesn't change its roughness. The thickness of the nanoring walls t varies from 110 to 160 nm for the all considered film thicknesses, and demonstrates its growth versus $D_{ring}$ and slight decrease versus d (Fig. 3e). The latter observation indicates that the amount of molten material, which is ejected during such nanojet formation and droplet ejection processes, increases with the increasing film thickness.

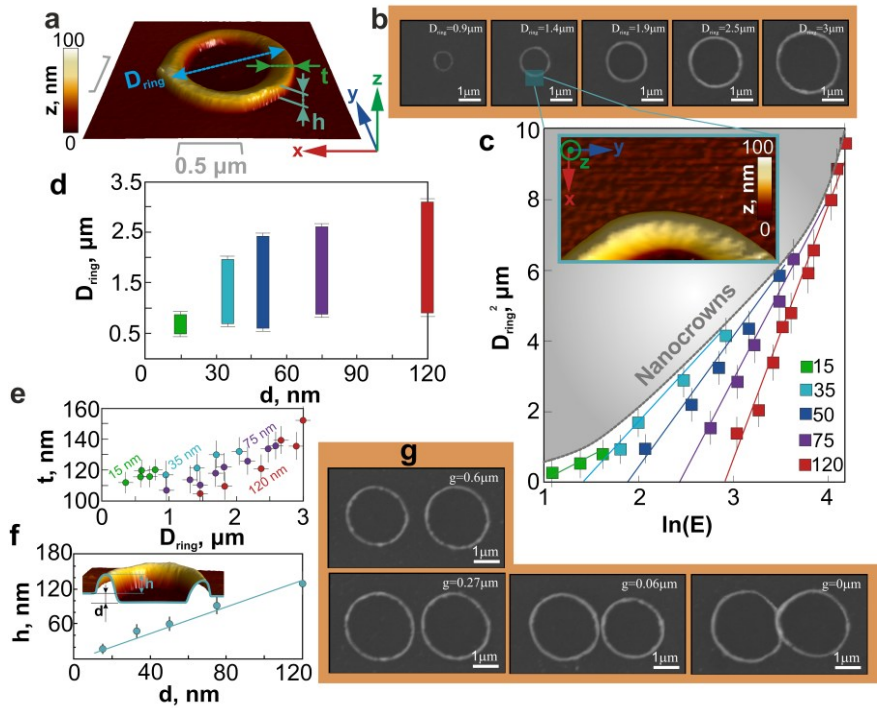

**Figure 3. Thickness- and energy-dependent dimensional parameters of nanorings.** (a) AFM image of a single Au nanoring, illustrating its main geometrical parameters. (b) SEM images of Au nanorings with the various diameters $D_{ring}$ fabricated in the 120-nm-thick Au film by single-pulse laser irradiation at variable energy E and subsequent $Ar^+$-ion beam polishing; (c) squared nanoring diameter $D_{ring}^2$ versus lnE (in nJ) for different values d; (d) variation of $D_{ring}$ as a function of d. (e) Nanoring thickness t versus its diameter $D_{ring}$ for different values d; (f) Nanoring height h as a function of d; (g) SEM images, showing nanoring dimers separated by different distances g.

Furthermore, the height h of the resolidified molten material measured above the initial film level (see the inset in Fig. 3f) determines the maximum height of the corresponding nanoring, emerging on the surface after the ion-beam polishing. The nanoring height measured using the AFM shows its linear growth in the range from 15 to 120 nm with the increasing film thickness (Fig. 3f).

Besides the demonstrated tunability of the three abovementioned main geometrical parameters of the nanorings – their diameter, wall height and thickness, there is another important experimental opportunity to manage gap width g between two adjacent nanorings (dimers). Properly tailored nanogap between two neighboring nanorings provides excitation of additional LPRs through plasmon-mode hybridization,[44] and strong magnetic response.[45] A series of SEM images in Fig. 3g demonstrates a sequence of nanoring dimers with variable gaps, which can be reduced down to their complete intersection, producing a complex plasmonic nanostructure.

Hence, the proposed two-stage nanofabrication technique enables to produce isolated plasmonic nanorings on the glass substrate with their broadly tailored dimensions: diameter – from 0.5 to 3.0 μm, wall thickness – from 110 to 160 nm and height – from 15 to 120 nm. Additionally, the proposed technique allows variation of the gap between two adjacent nanorings down to zero, yielding in complex-shaped nanopatterns of intersecting nanorings. In the next section, we demonstrate resonant plasmonic properties of the fabricated nanostructures and their sensing capabilities.

**Discussions**

In this section PL and Raman scattering of organic emitters deposited onto the abovementioned fabricated FPNs were investigated in order to reveal their localization and enhancement of electromagnetic fields in the visible spectral range through the corresponding LPR excitation. Figure 4a shows the DF scattering spectra measured from isolated Au nanorings with variable diameters under their unpolarized white-light illumination (see *Methods* section). All spectra demonstrate a broad scattering peak with its full width at a half-maximum ~100 nm, which is gradually red-shifted with the increasing nanoring diameter, apparently, due to the increased average thickness and the height of the nanoring[36]. This broad scattering peak indicates the excitation of the multiple LPR modes. Scattering spectra of an isolated Au nanoring ($D_{ring}$ = 1.4 µm, t = 100 nm, h = 50 nm) calculated for oblique plane-wave illumination using 3D FDTD simulations (see *Methods* section), demonstrate significant difference for p- and s-polarized light (the dashed curves in Fig. 4b) associated with the excitation of an intense dark SPP mode in the latter case [36,46]. Excitation of such modes by the s-polarized plane wave, obliquely illuminating the nanoring, is illustrated by a series of the calculated two-dimensional maps of the normalized squared electric field amplitude $|E|^2/|E_0^2|$ near the nanoring surface (Fig.4c-e).

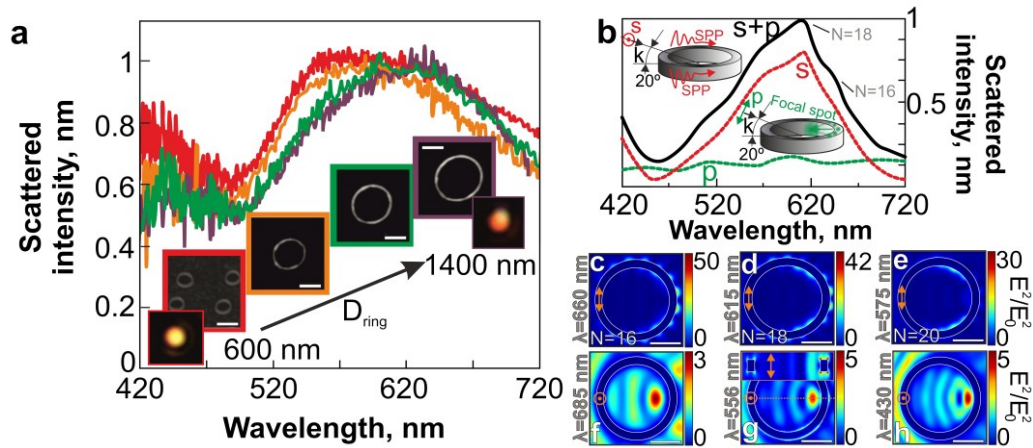

**Figure 4. Spectra of scattered light intensity and corresponding electrical near-fields for Au nanorings.** (a) Normalized DF back-scattering spectra measured from individual Au nanorings with their diameters $D_{ring}$=600, 900, 1150 and 1400 nm, obliquely illuminated (the incidence angle of 67°) by a non-polarized white-light source. The insets show the corresponding SEM and DF images of the nanorings (the scale bar corresponds to 700 nm). (b) Back-scattering spectra of a single Au nanoring ($D_{ring}$=1.4 µm, h=50 nm, t=100 nm) calculated for its s- and p-polarized plane-wave illumination (red and green dashed curves, respectively), as well as the total back-scattering spectrum for the both polarization directions (solid curve). The calculated two-dimensional maps of the normalized squared electric field amplitude $|E|^2/|E_0^2|$ near the nanoring surface for s- (c,d,e) and p-polarized (f,g,h) plane-wave illumination at the given wavelengths (details of the simulation geometry can be found in the *Supplementary information*). The scale bar corresponds to 500 nm and the polarization directions in all images are marked by the orange arrows.

As can be seen, the electric field near the nanoring represents a set of circumferentially spaced SPP-interference maxima, clearly demonstrating the processes of plasmon-polariton excitation on the left side of the nanoring by the s-polarized plane wave and their subsequent interference, which provides the standing SPP-wave distribution[39,46]. The resonant condition for such dark SPP modes is known to satisfy the following relationship: $\pi D_{ring}=N\cdot\lambda_{sp}/2$, where $\lambda_{sp}$ is the SPP wavelength, N=2,4,6, … [39,46]. Thus, for the given nanoring diameter $D_{ring}$=1.4 μm the main contribution to the DF scattering spectra in the visible spectral range is provided by three modes with N=16, 18 and 20, as illustrated in Fig. 4(c-e). Surprisingly, no excitation of dark SPP modes occurs under normal irradiation ($\theta^{in}$=0°) of the nanoring, while the intensity of such mode is maximal at $\theta^{in}$=90°, as was confirmed by the numerical calculation[46] and by previous experiments on angle-dependent PL enhancement from a self-organized Cd/Se quantum dot monolayer[47].

The scattering spectrum from the nanoring illuminated by the p-polarized plane wave demonstrates no significant features in the visible spectral range (Fig. 4b), indicating low efficiency of the SPP mode excitation, which is also supported by the calculated two-dimensional maps of the normalized squared electric field amplitude $|E|^2/|E_0|^2$ near the nanoring (Fig. 4f-h). As can be seen, the light intensity (the "focal spot") is maximal inside the inner surface[36] at the certain distance from the nanoring walls, rather than nearby, as for the case of the s-polarized illumination. The presence of such focal spot turns out to be attributed to constructive interference between the electric field of the radiation scattered on the right side of the nanoring and the electric field, representing the quadruple-like LPR mode (see inset in Fig.4g) excited in the nanoring by the p-polarized plane wave. The presented analysis of the dark SPP-modes excited under the s-polarized illumination is quite approximate owing to simultaneous excitation and cross-interaction of SPPs on the inner and outer nanoring sides, which perimeters will be differ owing to non-zero thickness of the nanoring walls. Nevertheless, the experimentally measured DF spectrum corresponds well to the total calculated spectrum (solid curve in the Fig.4b), with the low redshift of the experimental spectrum attributed to the well-known screening charge effect of the glass substrate[36]. Comparison between the obtained numerical and experimental results shows that some geometrical imperfections of the fabricated nanorings have no significant effect on their optical properties.

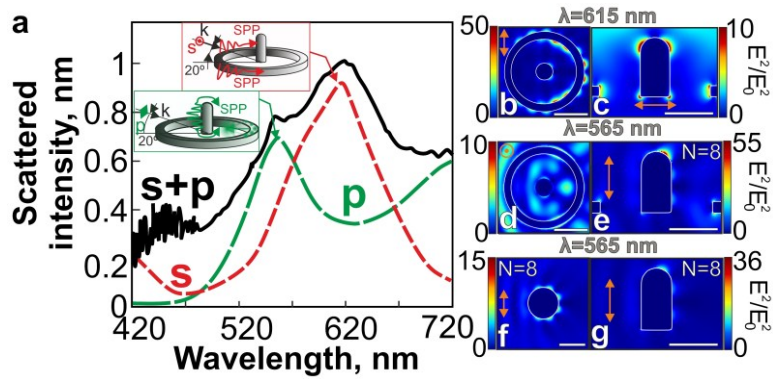

**Figure 5. Spectra of scattered light intensity and corresponding electrical near-fields for Au nanoring-nanorod ensembles.** (a) DF back-scattering spectrum (black solid curve) experimentally measured from the single isolated nanoring-nanorod structure ($D_{ring}$=1.4 μm, $h_{rod}$=600, $D_{rod}$=310 nm) under its oblique illumination by unpolarized white light, as well as corresponding calculated spectra for the s- and p-polarized illumination (dashed curves). The normalized squared electric field amplitude $|E|^2/|E_0^2|$ near the nanorod-nanoring ensemble for the s- (b,c) and p-polarized (d,e) plane-wave illumination at the given wavelengths. The normalized squared electric field amplitude $|E|^2/|E_0^2|$ near the isolated nanodisk with the equivalent perimeter (f) and the isolated nanorod (g). The scale bar corresponds to 500 nm and the polarization directions in all images are marked by the orange arrows and the circle.

DF back-scattering spectra measured from an Au nanorod-nanoring ensemble, containing both the nanorod (solid curve in Fig. 5a) and the nanoring, under unpolarized white-light illumination demonstrates an additional peak at 565 nm, which is not observed in the scattering spectrum of the single isolated nanoring of the same diameter. This peak can be associated with the excitation of the LPR modes in the isolated nanorod, if one takes into account the relatively large distance between the nanorod and inner walls of the nanoring, excluding the possibility of their mode hybridization[48]. This assumption is confirmed by the supporting FDTD calculations of the back-scattering spectra for the s- and p-polarized plane-wave illumination of the nanoring. Under the s-polarized illumination, the scattering spectrum (red dashed curve in Fig. 5a), as well as the E-field distribution near the nanorod-nanoring ensemble calculated at $\lambda=615$ nm (Fig. 5b), demonstrate no significant differences from the corresponding spectrum and the field distribution of the single nanoring (Fig. 4b,d). As seen, the most intense field is concentrated near its walls in the circumferentially spaced SPP interference maxima corresponding to the dark mode with N=18 (Fig. 5b), while the electric field intensity near the nanorod (Fig.5c) is significantly lower. The situation is different under the p-polarized plane-wave illumination of the nanorod-nanoring ensemble: the most intense electric field is concentrated near the nanorod (Fig.5d,e) in the dark LSP mode (N=8) similar to the that one, excited in the nanodisk[40] with perimeter $\pi D^{eq}_{disk}$ (Fig.5f), approximately equal to the perimeter of the vertical cross-section of the nanorod ~ $2h_{rod}+2D_{rod}$. However, unlike the equivalent nanodisk the asymmetric geometric shape of the nanorod leads to uneven intensity distribution with its maximum value reached near the nanorod top (Fig.5e,g). The excitation of such LPR mode apparently explains the appearance of the additional peak in the scattering spectra. It also should be noted that under the p-polarized illumination the nanoring, surrounding the nanorod, will act as a plasmonic lens, scattering the incident radiation back to the nanorod. Comparing the results of numerical calculations of the squared electric field near the isolated NR (Fig.5g) and the nanorod-nanoring ensemble (Fig.5e), we found 1.5-fold enhancement of the value $|E|^2/|E_0^2|$ in the latter case.

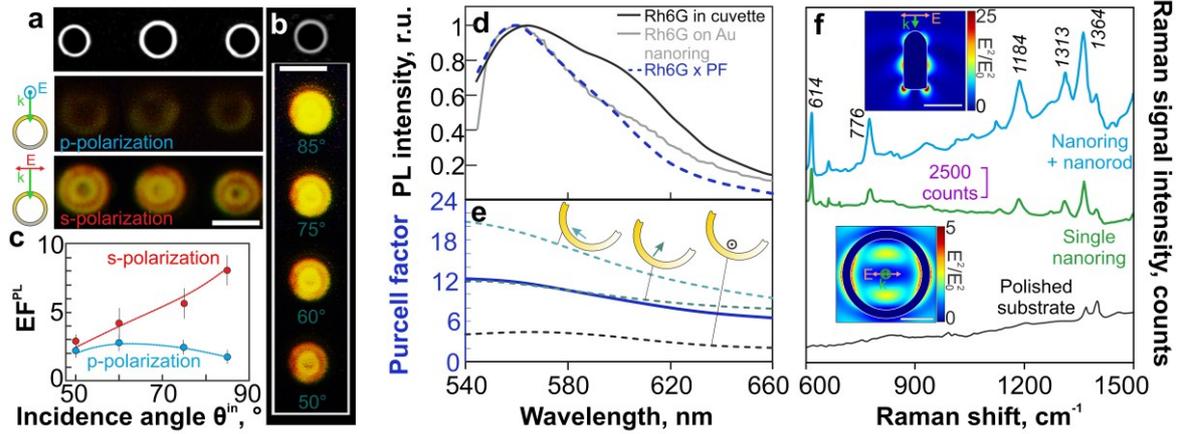

**Figure 6. Surface-enhanced PL and Raman scattering from fabricated FPNs.** (a) Reference SEM image of three Au nanorings ($D_{ring}$=1.2-1.45 μm) on the glass substrate and the corresponding PL images of a Rh6G-layer deposited on the nanorings. The PL images were obtained under oblique (angle of incidence $\theta^{in}$ =70° to the sample's normal) illumination by s- and p-polarized laser radiation; (b) PL images of the Rh6G-layer on the single Au nanoring obtained for different incidence angles $\theta^{in}$ in the range of 50°-85°. The color scale bar is the same for the all PL images. The scale bar in the SEM images in both the figures corresponds to 1 μm; (c) PL enhancement coefficient $EF^{PL}$ versus incidence angle $\theta^{in}$ of the pump laser radiation; (d) Normalized PL spectra of the Rh6G-layer measured from the ethanol solution of Rh6G in the quartz cuvette (black solid curve), the same spectrum multiplied by the average spectrally-dependent PF coefficient (dashed curve), and from the layer deposited on the Au NR (grey solid curve) shown in (b). (e) Calculated spectral dependence of the PF of the single nanoemitter placed near the Au nanoring (solid curve) averaged over the three orthogonal dipole orientations (dashed curves). (e) Raman spectra measured from a Rh6G layer on the single Au nanoring (green curve), nanorod-nanoring ensemble containing the nanoring with the same diameter and the nanorod (blue curve), as well as the glass substrate after complete removal of the Au film (black curve). The insets show the normalized squared E-field amplitude $|E|^2/|E_0^2|$ near the nanorod (top) and nanoring (bottom) calculated under normal plane-wave irradiation at the 633-nm wavelength (the scale bar corresponds to 500 nm).

In order to study resonant plasmonic properties of the fabricated nanostructures in more details, PL and surface-enhanced Raman scattering (SERS) signals were measured from a Rhodamine 6G (Rh6G) layer, deposited onto the fabricated FPNs. Figure 6a shows the reference SEM and PL images of the Rh6G layer, covering three Au nanorings with their diameters $D_{ring}$ between 1.2 and 1.45 μm, measured under oblique ($\theta^{in}$=70°) s- and p-polarized illumination at λ=532 nm. These PL images clearly indicate that the substantially enhanced signals from the Rh6G layer under the s-polarized illumination are apparently associated with intense electromagnetic near-fields near the nanoring walls, which are originating from the excitation of the dark SPP-modes. A series of PL images (Fig.6b) of the Rh6G layer on the single isolated Au nanoring obtained for the obliquely incident s-polarized exciting laser radiation with its incidence angle $\theta^{in}$, ranging from 50° to 85°, also supports this assumption. Under the p-polarized irradiation any substantial angle-dependent component of the PL signal were not observed (not shown). The maximal 8-fold PL enhancement (Fig. 6c) is expectedly achieved for the s-polarized illumination at $\theta^{in}$=85°, which is in good agreement with the enhancement factor (~ 7.5 times [47]) achieved for a chemically synthesized single-crystalline silver ring of almost ideal geometrical shape. The PL spectrum measured from the 10-nm-thick Rh6G layer deposited onto the Au nanoring demonstrates the obvious spectral narrowing, as compared with the spectrum measured from the Rh6G solution in ethanol (Fig. 6d), resulting from the PL decrease in the red spectral region.

It is well known that enhancement of spontaneous emission from an organic nanoemitter in the vicinity of a plasmonic nanostructure can be described by a Purcell factor (PF) coefficient[49]. To describe the experimentally observed PL enhancement as well as the emission spectrum reshaping of the Rh6G layer near the Au nanoring, we calculated a spectral PF dependence for a single dipole source placed 5 nm away from the nanoring wall ($D_{ring}$=1.2 μm). The PF value spectrally averaged from 532 nm to 670 nm, approaches 10, corresponding well to the experimentally measured maximal PL enhancement factor ($EF^{PL}$ ~ 8), which is apparently the upper limit for such nanostructures. The spectral PF dependence averaged over the orthogonal dipole orientations (dashed curves in Fig. 6e) also explains the narrowing of the Rh6G-layer emission spectrum on the Au nanoring. Indeed, the experimentally measured PL spectrum demonstrates good agreement with the PL spectrum of the Rh6G solution in the cuvette (black curve in Fig.6d) multiplied by the calculated PF spectral dependence (dashed curve in Fig. 6d), revealing plasmonic nature of the PL enhancement, rather that higher Rh6G molecule concentration near the nanoring walls.

SERS spectra of the Rh6G layer (Fig. 6f), covering the single isolated Au nanoring, the nanorod-nanoring ensemble, as well as the glass substrate after complete removal of the Au film demonstrate the average enhancement factor $EF^{SERS} \sim 10^3$ (for SERS measurement details and estimates see *Methods*). Higher SERS signal collected from the nanorod-nanoring ensemble is caused by excitation of a plasmon mode (N=6 at λ=633 nm, top inset in Fig. 6f), while for the Au nanoring only low-intense dipole-like oscillations can be excited at such irradiation conditions (bottom inset in Fig. 6f). Surprisingly, the Rh6G layer on the glass surface after complete removal of the Au film demonstrates appearance of two Raman bands – at 1313 cm$^{-1}$ and 1364 cm$^{-1}$, respectively, which are absent for the pure glass substrate (not shown). Such unexpected enhancement of the SERS signal on the dielectric substrate appears to be caused by small amount of metal conglomerates formed on the glass substrate through the metal film dewetting process. The heating of the film by Ar$^+$-ion beam may initiate this process when the film becomes very narrow, as confirmed by comparing the EDX spectra measured from the metal film on the glass substrate, the pure glass substrate and the glass substrate after complete removal of the film (For further information please see Fig.S3 and supplementary note 3 in the *Supplementary information*).

**Conclusion**

In summary, new simple two-stage technique was developed for fabrication of different complex-shape plasmonic nanostructures, including nanorods and nanorings, as well as nanorod-nanoring ensembles on glass substrates. In this technique, a noble metal film, covering a bulk glass substrate, is irradiated by a single tightly focused nanosecond laser pulse, accompanied by slow polishing of the laser-fabricated nanostructures by an accelerated argon ion ($Ar^+$) beam. Each nanosecond laser pulse locally modifies the initial metal film through initiation of fast melting and subsequent hydrodynamic processes, while the following $Ar^+$-ion polishing removes the rest of the film, revealing the hidden topography features and fabricating separate plasmonic structures on the glass substrate. We have demonstrated that the main geometric parameters of the fabricated NR and nanoring-nanorod ensembles are determined by the pulse energy and the metal film thickness, while the subsequent $Ar^+$-ion polishing allows height tuning for the nanostructures. Fabricated FPNs demonstrate both the 8-fold and up to $10^3$ plasmon-mediated enhancement of the PL and SERS signals from the Rh6G organic nanosized emitters, respectively. The plasmonic nature of this enhancement was rigorously examined and verified by direct measurements of the DF and PL spectra, as well as by angle-dependent character of the PL enhancement and supported by numerical calculations of corresponding electromagnetic near-fields and Purcell factor.

The reported two-step fabrication technique is quite simple providing the relatively high-repetition-rate fabrication of well-ordered arrays of different functional plasmonic nanostructures including nanorods, nanorings and nanorod-nanoring ensembles from thin films of main plasmonic noble-metal materials, once ultrastable nanosecond or femtosecond fiber lasers (pulse-to-pulse energy fluctuations below 0.5-% level) synchronized with scanning or micropositioning systems are used for such large-scale precise fabrication. Alternatively, single-pulse interference (3- or 4-beam) lithography [20,33], in which the pulse-to-pulse spontaneous fluctuations of the pulse energy is completely eliminated, turns out to be well applicable for fast high-repetition-rate fabrication of different types of nanostructures. Moreover, precise tailoring of the gap between two adjacent nanoring can be realized via spatial pre-shaping of laser pulses as it was recently demonstrated in [50]. Under these circumstances (use of ultrastable nano- or femtosecond fiber lasers, spatial beam modulation or multi-beam interference, synchronized laser-scanning/positioning hardware) the only local thin-film inhomogeneity (e.g., grain size, impurities, external contamination by dust particulates, etc.) and advanced vibration isolation appear to affect the large-scale nanofabrication quality. Regarding the ion-beam polishing/etching stage, its regimes are obviously yet to be optimized to enable nanometer-wide inter-element gaps in such arrays.

**Methods**

*Noble metal film fabrication.* Noble metal films (Au, Ag) with their thicknesses, ranging from 15 to 120 nm, deposited onto the optically flat bulk glass substrate by an e-beam evaporation procedure (Ferrotec EV M-6) at a pressure of $5 \cdot 10^{-6}$ bar and an average speed ~8 Å/c were used as samples for our laser nanostructuring experiments. To increase the adhesion of the deposited material to the glass substrate, the latter was pre-cleaned using a build-in $Ag^+$-ion source (KRI EH200). The film thickness was measured at the preliminary step by a calibrated piezoelectric resonator (Sycon STC-2002) mounted inside the vacuum chamber. These measurements were verified using an atomic force microscope (NanoDST, Pacific Nanotechnologies).

*Laser processing.* Laser nanostructuring of the noble metal (Au, Ag, Cu) films of variable thickness on the glass substrate was carried out using second-harmonic (532 nm) linearly-polarized pulses of a Nd:YAG laser (Quantel Ultra). The output energy pulse-to-pulse stability of the laser system used in our experiments was about 5%. The output radiation from the laser via the fiber coupler, equipped by a micropositioner and a focusing lens (NA = 0.25), was input into a segment of single-mode optical fiber (Thorlabs SM405) providing the radiation filtering for the resulting nearly Gaussian beam. This beam was then was focused onto the sample surface by a high-NA lens (x40, NA = 0.6) under complete filling of its input aperture providing the irradiation of the samples by laser pulses with a Gaussian–like focal spot with the characteristic radius $R_{opt} = 1.22\lambda \cdot (2NA)^{-1} \sim 0.34$ μm. The samples were arranged on a PC-driven three-dimensional motorized micropositioning stage (Newport XM series) with a minimal translation step of 50 nm along each axis and were moved from shot to shot. The pulse energy E was varied by means of a variable energy filter and measured by a sensitive pyroelectric photodetector (Coherent J-J-10SI-HE photodetector and Coherent EPM2000 Energy meter). For details, please see Fig. S1 in *Supplementary information*.

*$Ar^+$ beam polishing.* The nanostructures fabricated in the metal film surfaces after its single-pulse irradiation, were then polished by a normally-incident unfocused $Ar^+$ ion beam (the beam diameter ~6 mm, ion beam current – 30 μA, discharge current – 105 μA) till the depth of the initial film thickness, using a commercial apparatus (Hitachi IM4000). Average polishing rate were ranged from 0.5 to 1 nm/s depending on the chemical composition of the noble metal film. To avoid possible melting of the metal film and deformation of the resulting nanostructures under the ion beam, the polishing procedure was carried out in several successive cycles, where each cycle did not exceed 15 seconds, with metal film cooling for 5 min between such polishing cycles.

*Nanostructure characterization.* The resulting FPNs after each processing step were characterized by both low-vacuum scanning electron microscopy (SEM, Hitachi S3400N) and atomic force microscopy (NanoDST, Pacific Nanotechnologies) with ultra-thin cantilevers (NT MDT 01_DLC and NSC05_10°). Average deviations of the main geometric parameters of the fabricated nanostructures were generally caused by laser pulse energy fluctuations, mechanical vibrations of the experimental setup or some other factors (film quality, beam pointing instability etc.) and were systematically measured by their averaging over 50 repetitive experiments per each point presented in Figs. 2 and 3. The energy dispersive X-ray (EDX) spectroscopic microanalysis of the glass substrate (5x5 μm$^2$ area) after the complete removal of the metal film by the polishing $Ar^+$ beam was performed using the electron detector (ThermoScientific) at the acceleration voltage of 10 keV.

*Dark-field micro-spectroscopy.* To study the scattering FPN properties as a function of wavelength of the incident light, we used a DF microscopy scheme with independent excitation and collection optical channels[51]. White light (HL 2000 FHSA) was focused by an objective (Mitutoyo MPlanApo NIR 10, NA=0.26) onto the surface at an angle of incidence of 67°. The light scattered by each single FPN was collected by a second objective (Mitutoyo MPlanApo NIR 50, NA=0.42) and then was analyzed using a confocal spectrometer (HORIBA LabRam HR, cooled CCD Andor DU 420A-OE 325, 150 g/mm grating).

*Surface-enhance photoluminescence and Raman scattering measurements, evaluation of their enhancement factors.* The FPNs were covered by a 10-nm thick Rh6G layer. For PL measurements the Rh6G layer covering the fabricated nanostructures was obliquely irradiated by a CW semiconductor laser source (Milles Griot, the central wavelength λ=532 nm). Incidence angle $\theta^{in}$ was varied between 50° and 85° during the laser irradiation, being limited by the geometrical size and the working distance of the microscope lens (WD= 9 mm, NA=0.6) used to collect the PL signal. Laser polarization was switched between its s- and p-states by means of a λ/2-plate and a polarizer. The pump radiation scattered from the FPN and collected by the lens was completely blocked using a long-wavelength pass filter (with a band edge ~545 nm) placed in front of the CCD-camera used for visualization providing the pump wavelength attenuation ~$10^{-6}$. The enhancement coefficient $EF^{PL}$ was estimated as a ratio of the PL intensity $I_{sig}$ measured from the area containing the FPN to the intensity $I_{surf}$ from the Rh6G layer on the polished glass substrate area of the same size.

Surface-enhanced Raman scattering spectra were acquired using a micro-Raman spectrometer (HORIBA LabRam HR, AIST SmartSPM). Using a 2-mW, 632.8-nm HeNe laser, such spectra were recorded using a set of a 100× microscope objective (NA=0.9), a 600-g/mm diffraction grating and a thermoelectrically cooled charge-coupled device (CCD, Andor DU 420A-OE 325) array. Individual spectra were recorded from both single spots (0.86-μm diameter) on the substrates, and from a 1-cm thick cell of neat Rh6G for normalization. From these measurements, the SERS enhancement factor $EF^{SERS}$ averaged over the excitation beam size, was estimated as $EF^{SERS} = (I_{SERS}/I_{norm})(N_{norm}/N_{SERS})$, where $I_{SERS}$, $I_{norm}$ and $N_{SERS}$, $N_{norm}$, are the intensities and the numbers of the probed molecules on the sample and on the reference surface at specific Raman band (1364 cm$^{-1}$), respectively. $N_{norm}$ was evaluated using the molecular weight and density of R6G, and the effective interaction volume of the Gaussian laser beam in the sample. We calculated the effective excitation volume as $V_{ex} = \pi r^2 H$, where $r$ is the radius of the beam ($r \approx 0.43$ μm) and $H$ is the thickness of optical section ($H \approx 11.7$ μm). Thus, we estimated the effective excitation volume of 6.8 μm$^3$ for our Raman microscopy with 633-nm excitation using the objective. Then, $N_{norm}$ was calculated using the expression $N_{norm} = CN_A V_{ex} \rho/M = 3.63 \cdot 10^9$ molecules, where $C$ is the relative concentration (3%), $\rho$ is the density of R6G (1.26 g/cm$^3$), $M$ is the molar mass of Rh6G (479 g/mol) and $N_A$ is the Avogadro constant (6.02·10$^{23}$ mol$^{-1}$). Likewise, the number of the dye molecules $N_{SERS}$ within the laser beam focused onto 10-nm thick R6G layer over the nanostructure can be estimated as $N_{SERS} = CN_A \pi r^2 h \rho/M = 3.1 \cdot 10^6$ molecules.

*Numerical simulations.* We calculated back-scattering spectra and electric field distributions near the Au nanorings and nanorod-nanoring ensembles using finite-difference time-domain simulations (Lumerical Solutions package[52]). The dielectric function of Au was modeled using experimental data from[53]. Total-field scattered-field source with the wavelength ranging from 300 to 900 nm was used to irradiate the FPNs. The size of the square unit cell was as small as 1x1x1 nm$^3$, and the computational volume was limited by perfectly matched layers. Rounded corners (radius of curvature = 10 nm) were included to the sharp edges of the nanorods and nanorings to avoid staircase effect, which can cause significant E-field amplification. (For details of the modeling geometry please see Fig. S2 in the *Supplementary Information*).

The Purcell effect is defined as a variation of a spontaneous emission lifetime of a quantum source induced by its interaction with an environment. Specifically, PF was considered as an enhancement factor for the power emitted by the emitter in the inhomogeneous medium in comparison with the power emitted by the same emitter in free space. The radiated power (P) of a dipole (emitter) is described by its radiation resistance ($R_{rad}$) as $P = (j\omega d/l)^2 R_{rad}$, where $j=\sqrt{-1}$, $\omega$ is the radial frequency, d is the dipole moment, l is the effective length of the dipole. Therefore, the PF can be written in the form $PF = P/P_0 = R_{rad}/R_{0,rad}$, where index "0" corresponds to the emitter in free space[54]. In the commercial software package (CST Microwave Studio), such point dipole is modeled as an optically very short dipole ($l \ll \lambda$) of a perfectly conducting wire excited by an ideal current source. Thus, if the emitter is low-loss one, we can write an equivalent relation for the PF of the nanostructure for a low-loss emitter: $F = Re(Z_{in})/Re(Z_{0,in})$, where $Z_{in}$ and $Z_{in,0}$ are the input impedances of the dipole nearby and infinitely far from the nanostructure, respectively, and calculate the values of the impedances numerically[54]. The numerical calculation of the impedances was performed for such dipoles placed at the distance of 5 nm from the surface of gold nanostructures, in agreement with Fig. 7b.

**Supplementary Information**

Schematic of the experimental setup, FDTD calculation settings, EDX spectra and the resulting elemental compositions. Supplementary information accompanies this paper at

http://www.nature.com/scientificreports

**Acknowledgements**

Authors from IACP and FEFU are grateful for partial support to the Russian Foundation for Basic Research (Projects nos. 14-02-31323-mol_a, 15-02-03173-a, 15-02-08810-a). The project was also financially supported by the Russia Federation Ministry of Science and Education, Contract № 02.G25.31.0116 of 14.08.2014 between Open Joint Stock Company "Ship Repair Center "Dalzavod" and RF Ministry of Science and Education. A.A.

Kuchmizhak is acknowledging for partial support from RF Ministry of Science and Education (Contract No. MK-3056.2015.2) through the Grant of RF President. This work was also supported by the Government of the Russian Federation (Grant 074-U01) through ITMO Visiting Professorship Program (for S.I. Kudryashov). The modeling of fluorescence enhancement and Raman measurements were financially supported by Russian Science Foundation (Grant 15-19-00172).

**Author Contributions**

A.K. and S.G. performed laser processing experiments and FDTD simulations. S.M. performed the Purcell factor calculations. V.M. carried out the dark-field and SERS measurements. S.G. carried out photoluminescent measurements. Yu.K., O.V. and S.K. contributed to the analysis and discussion of the data and edited the manuscript. S.K., S.M. and A.K. wrote the manuscript.

**Additional Information**

Competing financial interests: The authors declare no competing financial interests.

# Supplementary Information

**Ion-beam assisted laser fabrication of sensing plasmonic nanostructures**
Aleksandr Kuchmizhak,[*] Stanislav Gurbatov, Oleg Vitrik, Yuri Kulchin, Valentin Milichko, Sergey Makarov, and Sergey Kudryashov

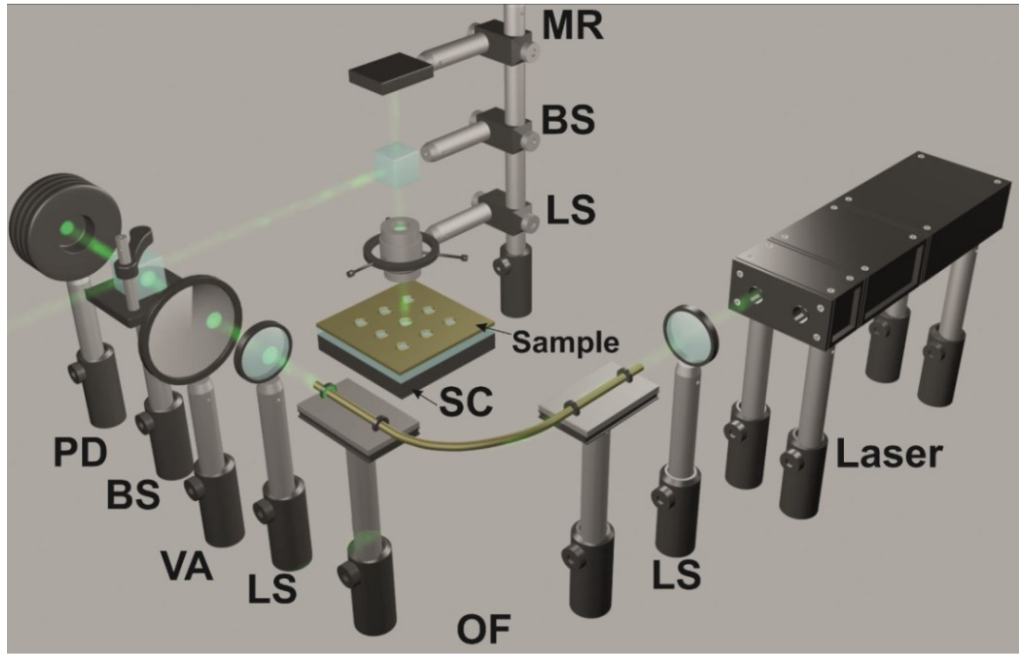

**Fig.S1. Schematic of the experimental setup:** LS- lens, MR – mirror, BS – beam splitter, PD – photodetector, OF – optical fiber, SC – scanning platform, VA – variable energy attenuator.

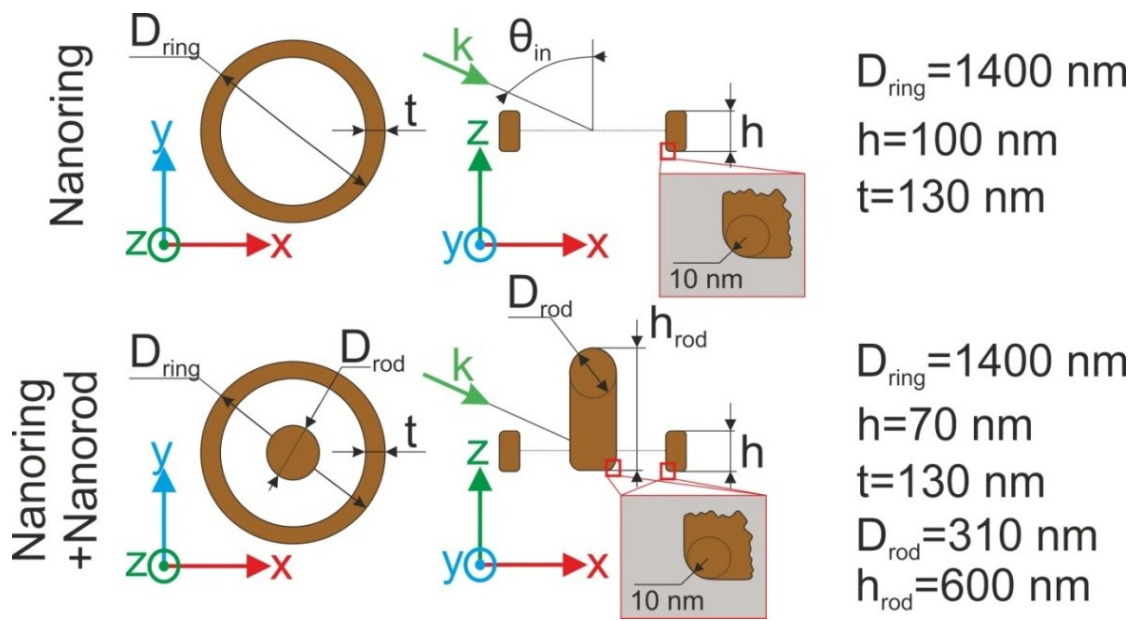

**Fig.S2. The details of the FDTD simulations of the nanorings and nanoring-nanorod ensembles.**

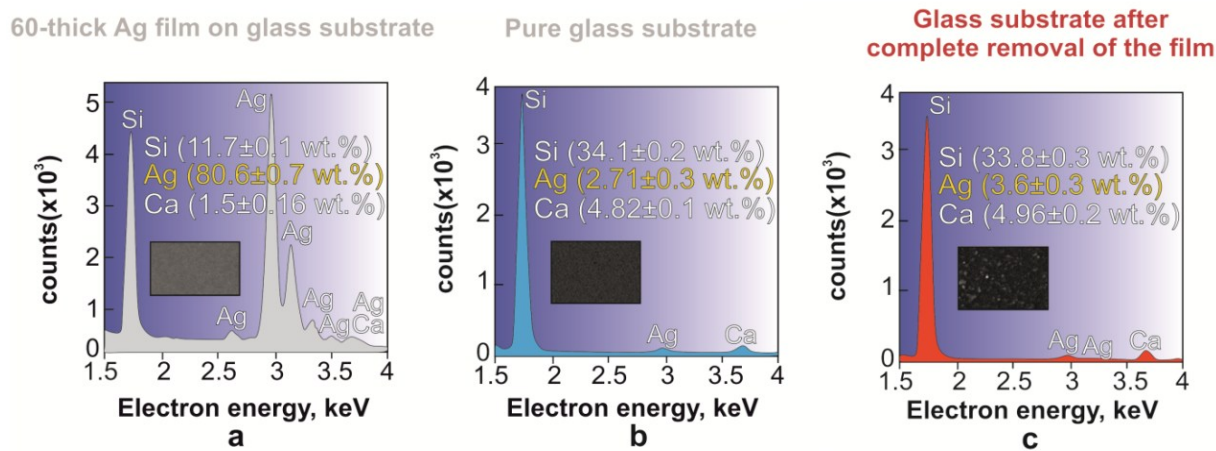

**Fig.S3. EDX spectra** measured for the 60-nm-thick Ag film on the glass substrate (a), the pure glass substrate (b) and for the glass substrate after complete removal of the Ag film (c). Insets show SEM images of the corresponding areas (5-μm wide), where the data acquisition was performed.

**Supplementary note.**

To ensure that the entire metal film was completely removed from the glass substrate after $Ar^+$-beam polishing procedure, we have acquired and compared EDX spectra from the 60-nm-thick Ag film on the glass substrate, pure glass substrate as well as from the polished glass surface near the fabricated plasmonic nanostructures (Fig.S3). As seen, the relative Ag-content from the polished surface is slightly (less than 1 wt.%) higher, than the reference noise level measured for the pure glass substrate, and is significantly smaller, than the Ag-content for the metal film, indicating that very small amount of silver remained on/inside the glass substrate. We believe that the residual separate small Ag nanocrystallites won't form a continuous film, thus not influencing significantly on the properties of the fabricated isolated functional plasmonic nanostructures. Note that, EDX spectra acquired for the Au and Cu nanostructures, demonstrated similar values of the corresponding metal content on the polished glass substrates after the complete removal of these metal films.